# Effects of transition metals on physical properties of M$_2$BC (M = V, Nb, Mo and Ta): a DFT calculation

P. Barua[1], M. M. Hossain[1*], M. A. Ali[1], M. M. Uddin[1], S. H. Naqib[2], and A. K. M. A. Islam[2,3]

[1]Department of Physics, Chittagong University of Engineering and Technology, Chittagong-4349, Bangladesh
[2]Department of Physics, University of Rajshahi, Rajshahi 6205, Bangladesh
[3]Department of Electrical and Electronic Engineering, International Islamic University Chittagong, Kumira, Chittagong, 4318, Bangladesh

*Corresponding author, email: mukter.phy@gmail.com

## Abstract

In the present study, the effects of transition metals on structural, electronic, elastic, optical and thermodynamic properties of M$_2$BC (M = V, Nb, Mo and Ta) have been investigated using the density functional theory based first-principles method. The electronic band structures along with Fermi surface, elastic anisotropy, Vickers hardness, analysis of Mulliken populations, optical and thermodynamic properties are studied for the first time. The optimized unit cell parameters are compared with available theoretical and experimental results and a reasonable agreement is recorded. The mechanical stability of these compounds is confirmed by the calculations of single crystals elastic constants using the Born criteria conditions. The compounds herein exhibit metallic conductivity where major contribution comes from the *d*-orbital electrons. The total density of states at $E_F$ are found to be 9.15, 6.77, 6.37 and 5.83 states/eV/unit cells for M$_2$BC (M = V, Nb, Mo and Ta) compounds, respectively. The hardness values of 10.71, 12.44, 8.52 and 16.80 GPa are noted for the M$_2$BC (M = V, Nb, Mo and Ta) compounds, respectively. The value of bulk modulus, *B* is found to increase in the sequence of $B$ (V$_2$BC) < $B$ (Nb$_2$BC) < $B$ (Ta$_2$BC) < $B$ (Mo$_2$BC), indicating Mo$_2$BC is highly stiff among the compounds under study. The Mo$_2$BC and Ta$_2$BC compounds might be considered as potential candidates for cutting and forming tools due to the moderately ductile and highly stiff behavior compared to other benchmark hard coating materials such as TiN, TiAlN, Ti$_{0.5}$Al$_{0.5}$N and c-BN. Ta$_2$BC compound could also be a promising thermal barrier coating (TBC) material due to its low thermal conductivity, Debye temperature and damage tolerant behavior. Different anisotropy calculations (shear anisotropic factors, percentage anisotropy factors and universal anisotropic index) confirm that the compounds are structurally anisotropic in nature. The bond between C-V/Nb/Mo/Ta shows that the degree of covalency is higher compared to B-V/Nb/Mo/Ta bond. Various optical functions (such as dielectric constants, refractive index, photo-conductivity, absorption, loss function and reflectivity) are calculated and discussed using established formalism. The amount of reflectivity is always more than 50% with no significant change in the near infrared, visible and near ultraviolet region (upto ~ 6 eV), this result makes the compounds promising coating materials to diminish solar heating.

*Keywords*: metallic boro-carbides, Band structure, Optical properties, Thermodynamic properties

## 1. Introduction

Many ceramic materials like TiAlN, TiN, $Al_2O_3$, c-BN, $Ti_{0.5}Al_{0.5}N$, $Ti_{0.25}Al_{0.75}N$ and $Cr_{0.5}Al_{0.5}N$ are usually used for surface coating on tools, dies, and many mechanical parts due to their interesting properties such as high hardness, stiffness, and chemical stability to enhance their lifetime and performance [1-5]. Hard and wear-resistant coating materials as ceramics often develop cracks and subsequently their propagation; thereby the tool lifetime and performance are diminished. In order to reduce crack initiation and growth, it would be reasonable to investigate new materials which render high stiffness with moderate ductility. Intrinsic ductility of the materials usually depends on the ratio of the bulk and shear modulus and a positive Cauchy pressure value [6]. Nanolaminates compounds so called MAX phases (M: transition metal; A: group 13–15 A element; X: C or N) exhibit unusual properties like low density, high elastic stiffness, refractory and resistant to high-temperature oxidation [7]. The $Mo_2BC$ belongs to orthorhombic structure with space group *Cmcm* and is also known as nanolaminated since it is composed of nanolayers. This structure can be explained as a combination of boride and carbide sub-cell. The B atom forming zigzag chains, is located in the $Mo_6B$ trigonal prisms and the C atoms are at $Mo_6C$ octrahedral site [8, 9].

The nanolaminate $Mo_2BC$ was successfully synthesized as a thin film using magnetron sputtering at substrate temperature 900° C and showed high stiffness, moderate ductility and chemical stability even at higher temperatures [2]. The superconducting properties of the $Mo_{2-x}M_xBC$ boro-carbides (M = Zr, Nb, Rh, Hf, Ta, W) were also reported. Single crystals of pure $Mo_2BC$ of nearly 10 mm long and 3 mm in diameter were grown by a Czochralski method [9]. Due to the promising prospect of industrial applications, a systematic theoretical investigation was required on the $Mo_2BC$ to identify the new materials which show the aforementioned properties. To do this, the substitutional effects of Mo atom by other transition metals such as Ti, V, Zr, Nb, Hf, Ta and W were studied using *ab-initio* calculations. The stability of all the compounds has been reported by the calculation of formation energy as well as single crystal elastic constants (mechanical stability) [6].Limited number of experimental and theoretical studies only focused on the structural forms and mechanical properties of $M_2BC$ (M = Mo, Ti, V, Zr, Nb, Hf, Ta and W) [2, 6,10-12]. As far as electronic properties are concerned, only the energy density of states (DOS) and charge density of $Mo_2BC$ and $Zr_2BC$ were studied [2, 6]. To the best of our knowledge, still many important physical properties such as Fermi surface topology, optical constants, Vickers hardness, Mulliken population analysis, charge density distribution, and thermodynamic properties are yet to be investigated. This provides us with a strong motivation to study the theoretical properties of $M_2BC$ in detail in this paper. It is expected that the $Mo_2BC$ compound and their counterparts could show more desirable properties in some cases, compared to the widely studied MAX phase compounds.

The optical properties are of great importance in giving insight into the fundamental electronic properties and potential optoelectronic applications. The hardness of a material as well as charge density are very important for application point of view that can be understood by the analysis of Mulliken population. It is also well known that the study of thermodynamic properties such as Debye temperature, specific heats, thermal expansion coefficient etc. are extremely important in solid state science and various industrial applications because these properties reveal particular behavior under high temperature and pressure conditions. Mechanical, thermoelectric and optical properties are closely related to the electronic properties of a material. Minor variations in the electronic structure may cause strong changes in these properties [13].

In the present study, we will pay attention to the theoretical results of electronic, optical, analysis of Mullikin population, theoretical Vicker hardness and thermodynamic properties along with Fermi surface and charge density of isostructural $M_2BC$ (M = V, Nb, Mo and Ta) for the first time in order to understand these compounds further. In addition, DOS and elastic properties will also be discussed in order to provide some additional information on these compounds for their effective use in device applications.

**2. Computational methods**

Density functional theory (DFT) calculations, in which the ground state of the system is found by solving the Kohn-Sham equation [14] are performed within generalized gradient approximation (GGA) for electron exchange correlation, using the CAmbridge Serial Total Energy Package

(CASTEP) [15]. The wave functions are expanded with plane waves, and ultrasoft pseudo-potential is used to reduce the number of plane waves [16, 17]. The cut-off energy for the plane wave expansion is 500 eV for all the compounds. The GGA is used for the exchange-correlation term, and the functional form is of Perdew-Burke-Ernzerhof (PBE) type [18]. Density mixing is used to the electronic structure and Broyden Fletcher Goldfarb Shanno (BFGS) geometry optimization is used to optimize the atomic configuration [19]. Periodic boundary conditions are used to calculate the total energies of each cell. k-point sampling for these compounds is carried out at 8×2×8 special points in a Monkhorst-pack grid [20]. Geometry optimization is performed using convergence thresholds of $5 \times 10^{-6}$ eV atom$^{-1}$ for the total energy, 0.01 eV Å$^{-1}$ for the maximum force, 0.02 GPa for maximum stress and $5 \times 10^{-4}$ Å for maximum displacement. Elastic constants were calculated by the 'stress-strain' method within the CASTEP program. The bulk modulus, $B$ and shear modulus, $G$ were obtained from the calculated elastic constants, $C_{ij}$. The Vickers hardenss ($Hv$) was calculated by using the Mullikin population analysis.

## 3. Results and Discussion
### 3.1 Structural properties

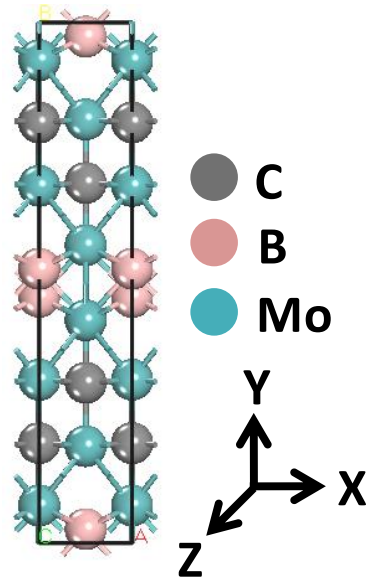

**Fig. 1.** Optimized crystal structure (two dimensional) of Mo$_2$BC.

As the first step, the equilibrium lattice parameters for M$_2$BC (M = Mo, V, Nb and Ta) compounds were obtained by optimizing the unit cell as a function of energy of the unit cell. The Wyckoff positions of the atoms are as follows: M(1) and M(2) atoms are located at (0,

0.0721, 0.25) and (0, 0.3139, 0.25), respectively; B atoms are at (0, 0.4731, 0.25) and C atoms at (0, 0.1920, 0.25). Fig. 1 shows the unit cell structure of $Mo_2BC$. The total energy of pseudo-atomic calculation was investigated for every compound. For instance, pseudo-atomic calculation performed for B $2s^2\ 2p^1$, C $2s^2\ 2p^2$ and Mo $4d^5\ 5s^1$ converged in 17, 19, and 22 iterations to a total energy of -70.4147, -145.6544 and -209.9349 eV, respectively. The crystal structures of these compounds are orthorhombic in which the unit cell contains four formula units. The calculated and experimental values of lattice constants of these compounds are summarized in Table 1.The calculated lattice parameters for the compounds are well justified with other theoretical and experimental data as the discrepancies in these parameters were less than 1%.

**Table 1:** Optimized lattice parameters, $a$, $b$ and $c$ (Å), unit cell volume $V$ (Å$^3$) of $M_2BC$ (M = Mo, V, Nb and Ta) compounds.

| Optimized parameters | $a$ | $b$ | $c$ | $V$ | Ref. |
|---|---|---|---|---|---|
| $V_2BC$ | 2.998 | 16.647 | 2.948 | 147.17 | This |
|  | 2.999 | 16.770 | 2.956 | 148.70 | [6] |
| $Nb_2BC$ | 3.224 | 18.239 | 3.141 | 184.71 | This |
|  | 3.227 | 18.345 | 3.149 | 186.41 | [6] |
| $Mo_2BC$ | 3.137 | 17.702 | 3.094 | 171.86 | This |
|  | 3.094 | 17.768 | 3.091 | 169.93 | [2] |
|  | 3.119 | 17.580 | 3.082 | 169.01 | [6] |
|  | 3.086 | 17.350 | 3.047 | 163.14 | [9] |
|  | 3.086 | 17.350 | 3.047 | 163.14 | [21] |
| $Ta_2BC$ | 3.267 | 18.445 | 3.180 | 191.69 | This |
|  | 3.221 | 18.246 | 3.140 | 184.51 | [6] |

### 3.2 Electronic properties

The electronic properties of a compound can be obtained from the band structure, partial density of states (PDOS) and total density of state (TDOS) which are also closely related to charge density distribution and the Fermi surface. Moreover, these properties are very convenient to elucidate bonding nature and other relevant properties of materials. The results of band structure calculations, for the first time, along the highly symmetric directions within the *K*-space for

$V_2BC$, $Nb_2BC$, $Mo_2BC$ and $Ta_2BC$ are presented in Fig. 2 (a)-(d), where the horizontal dashed line is shown as the Fermi level, $E_F$. The band belongs to valence and conduction bands are indicated with black line where the bands crossing the Fermi level ($E_F$) are presented by different colors. In the band diagrams, several dispersive bands crossed the $E_F$ especially along the G-Z direction for all the samples exhibiting no existence of band gap at $E_F$.

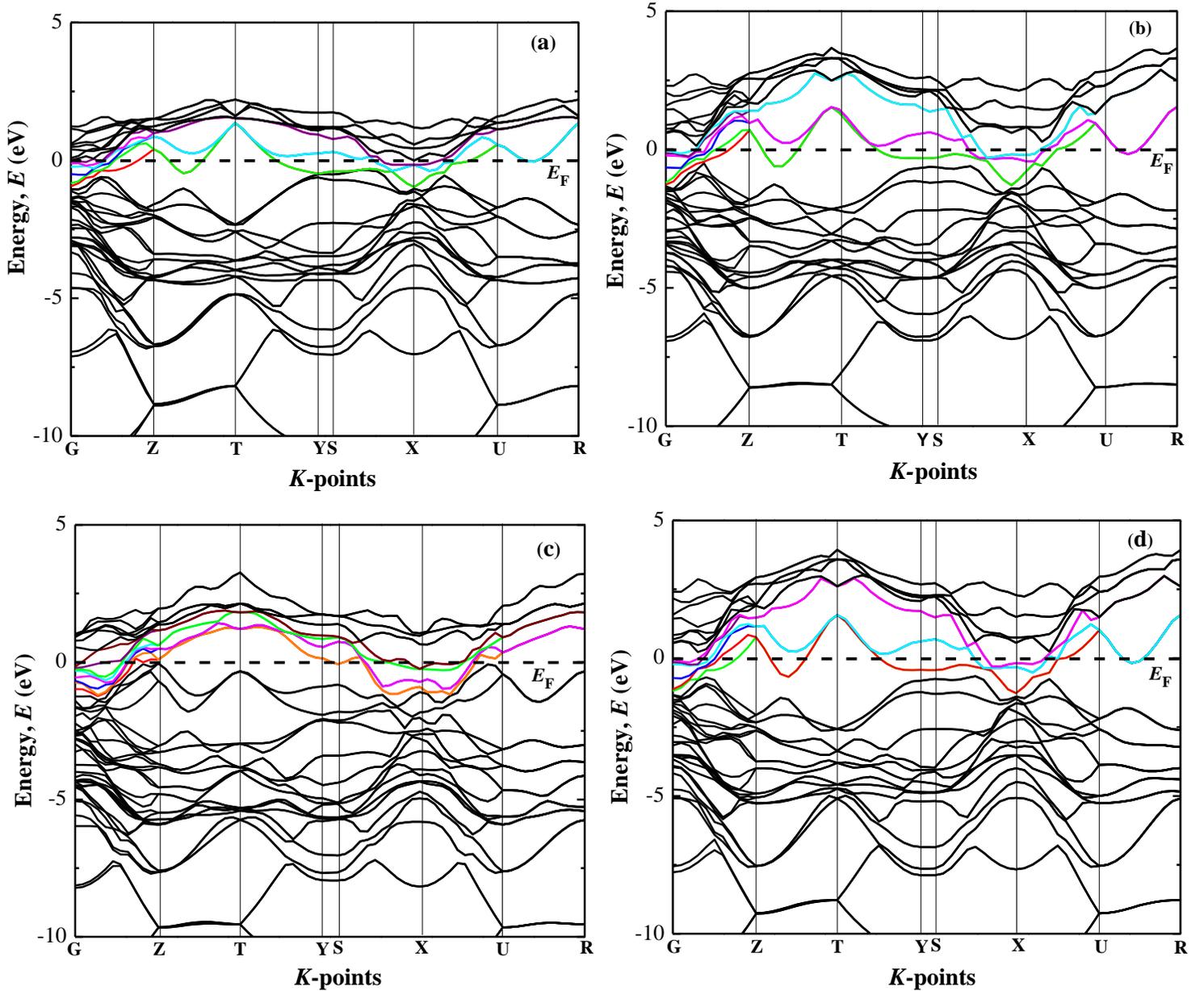

**Fig. 2.** Calculated electronic band structure of (a) $V_2BC$ (b) $Nb_2BC$ (c) $Mo_2BC$, and (d) $Ta_2BC$ along the high symmetry directions in the Brillouin zone at ambient conditions.

Therefore, all the compounds herein must exhibit metallic conductivity. In addition, we see nearly flat and broader bands along Y-S direction in the vicinity of $E_F$ for all compounds which is an indication of large DOS value. Importantly, the overall band profiles of the calculated compounds in the present research are nearly analogous to those noticed in previous articles [2, 6, 22-26].

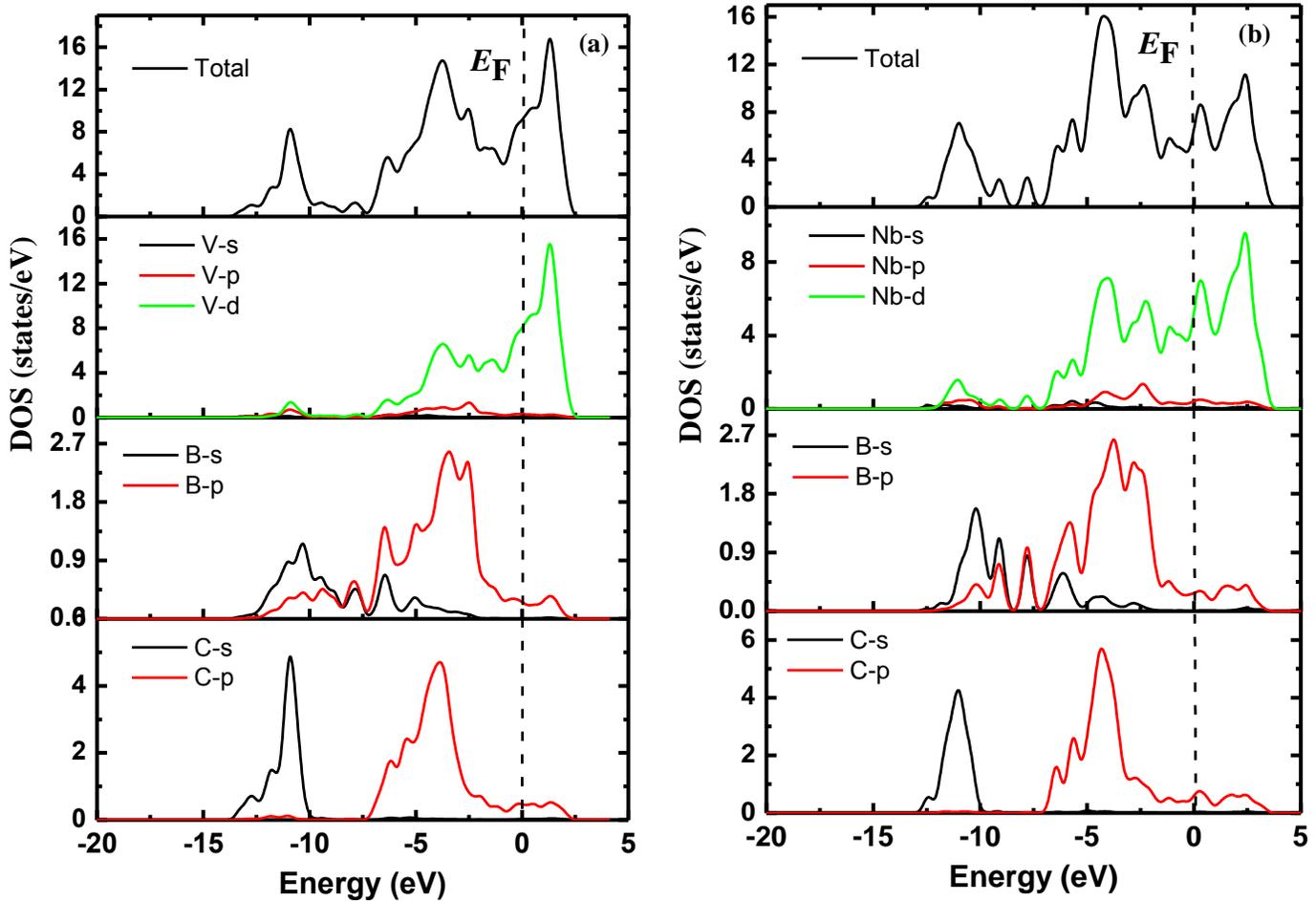

**Fig. 3.** Total and partial electron energy density of states of (a) $V_2BC$ and (b) $Nb_2BC$.

To estimate contributions of different orbitals/atoms and nature of chemical bonding in $M_2BC$ (M = V, Nb, Mo and Ta) compounds, TDOS and PDOS are calculated and is depicted in Fig. 3(a)-(b) and 4(c)-(d), respectively. The sharp peak of the DOS at the $E_F$ is a symbol of structural instability, whereas a deep and broad valley of DOS might be responsible for the structural stability. Therefore, the overall DOS profile of these compounds reveals structural stability. However, Figs. 3 and 4 show a strong hybridization among the B-2$p$, C-2$p$ and V-3$d$/ Nb-4$d$/Mo-4$d$/ Ta-5$d$ orbitals for the compounds. It is observed that the TDOS around the $E_F$ arises mostly from V-3$d$ ($V_2BC$), Nb-4$d$ ($Nb_2BC$), Mo-4$d$ ($Mo_2BC$) and Ta-5$d$ ($Ta_2BC$) orbitals with a small contributions of B-2$p$ and C-2$p$. Therefore, electronic charge should transfer from 3$d$ ($V_2BC$)),

4$d$ (Nb$_2$BC), 4$d$ (Mo$_2$BC) and 5$d$ (Ta$_2$BC) to 2$p$ (B and C), respectively owing to the strong interactions among them. These strong hybridizations imply strong ternary covalent B-C-V, B-C-Nb, B-C-Mo and B-C-Ta bonds in V$_2$BC, Nb$_2$BC, Mo$_2$BC and Ta$_2$BC compounds, respectively. Notably, the value of TDOS at $E_F$ is found to be 9.15, 6.77, 6.37 and 5.83 states/eV/unit cells for V$_2$BC, Nb$_2$BC, Mo$_2$BC and Ta$_2$BC compounds, respectively. The highest TDOS is obtained (9.15 states/eV/unit cell) at $E_F$ for the V$_2$BC compound. The calculated DOS values for the compounds are in good agreement with previously reported values [2, 6].

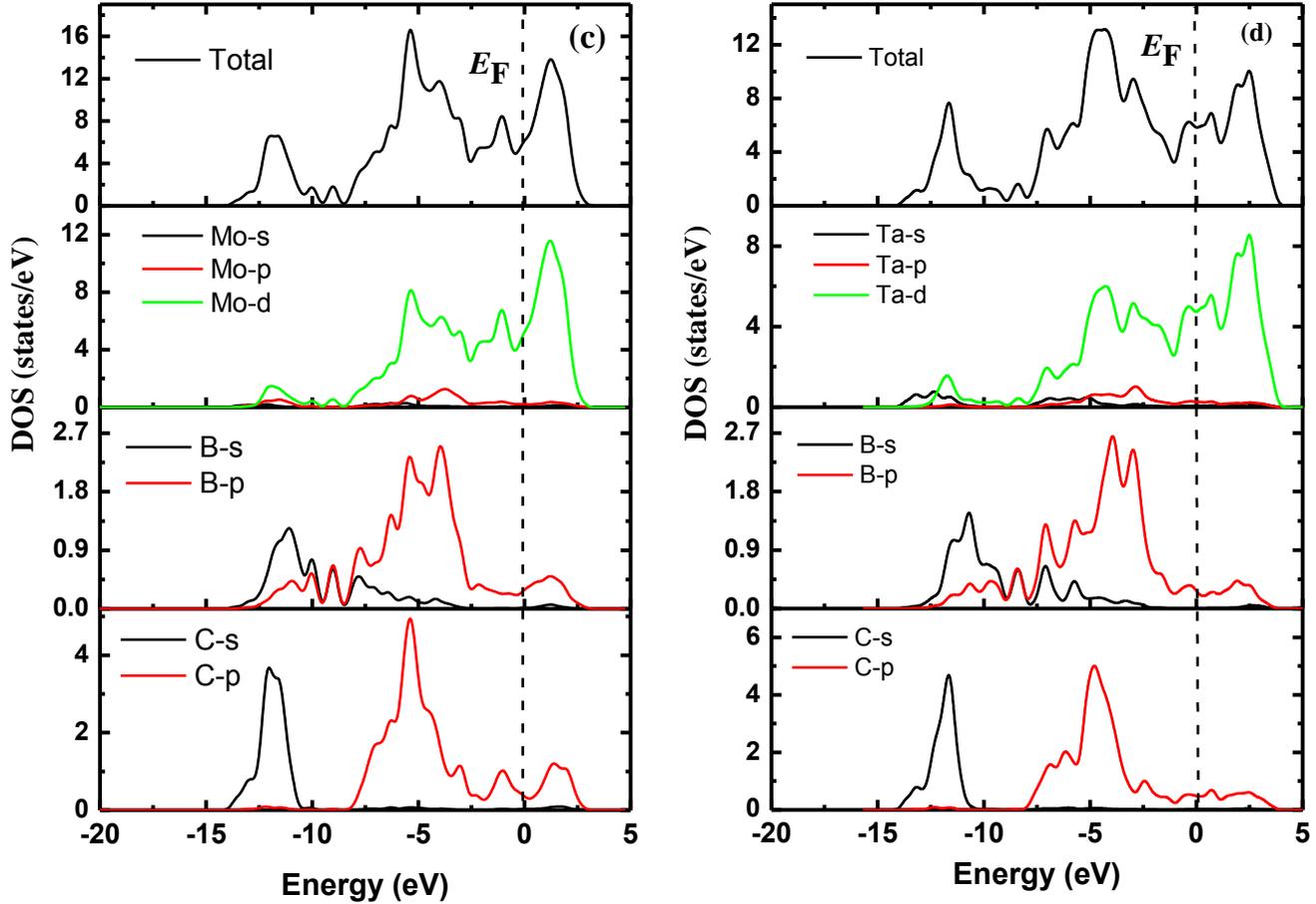

**Fig. 4.** Total and partial electronic energy density of states of (c) Mo$_2$BC and (d) Ta$_2$BC.

### 3.3 Charge density and Fermi surface

The charge density distribution in (110) planes of M$_2$BC (M = V, Nb, Mo and Ta) compounds has been studied to make sense of the bonding properties as illustrated in Fig. 5. An atom of high electronegativity always pulls the electron density towards itself [27]. It is noteworthy that strong accumulation of electronic charges around C and V/Nb/Mo/Ta is greater than those around B and V/Nb/Mo/Ta due to the large variation in electronagitivity between these two atoms. Therefore, the directional V/Nb/Mo/Ta-C bond is more covalent in nature than the V/Nb/Mo/Ta-B bond. Strong hybridization between V/Nb/Mo/Ta($d$) state and B/C($p$) state at the $E_F$ is mainly responsible for making this bond. In addition, the non-directional characteristics of the bonding (metallic bonding) between V-V, Nb-Nb or Ta-Ta transition metals could be formed. These results are also consistent with our Mulliken atomic population analysis (Table 5) and DOS (Fig. 3, 4). The chemical bonding nature of these compounds are somewhat similar to the bonding nature of MAX phase nanolaminates [28, 29]. Our results reveal that the $M_2BC$ compounds possess mixed (covalent, ionic and metallic) chemical bond characteristics.

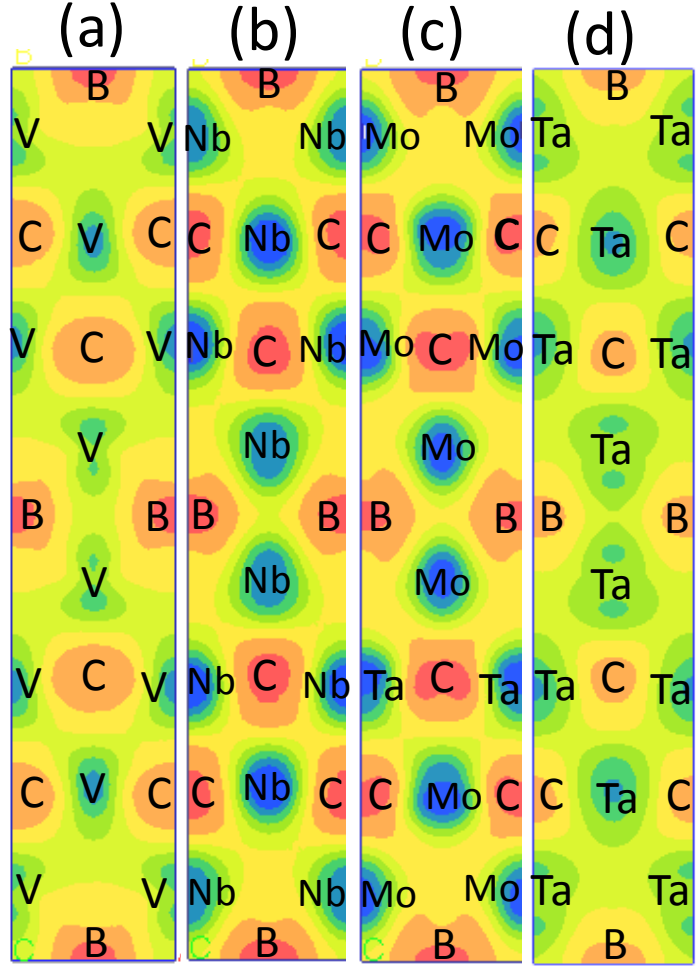

**Fig. 5:** Charge density distribution of (a) $V_2BC$, (b) $Nb_2BC$, (c) $Mo_2BC$ and (d) $Ta_2BC$ compounds in the (110) plane.

The Fermi surfaces of $M_2BC$ (M = V, Nb, Mo and Ta) compounds are depicted in Fig. 6 where solely the bands crossing the $E_F$ are shown. The Fermi surface topologies of $V_2BC$, $Nb_2BC$ and $Ta_2BC$ compounds are almost similar but different in case of $Mo_2BC$. For $V_2BC$, $Nb_2BC$ and $Ta_2BC$ compounds, three electron-like sheets with cylindrical cross sections are noticed to be centered along the X-G direction. Two similar sheets with little curved plane are positioned at the opposite directions of these three sheets. Two sheets with curved plane are also there along G-Z direction. The surfaces also include of some hole-like sheets centered at the corners of the Brillouin zone. For $Mo_2BC$, five electron-like sheets with hexagonal cross-sections are centered along X-G direction. Two similar sheets, far from five electron-like sheets are positioned at the opposite directions. The Fermi surfaces of these four compounds are formed mainly by the low-dispersive $4d/5p$ and $2p$ orbitals, which should be mainly responsible for the electrical conductivity of these compounds.

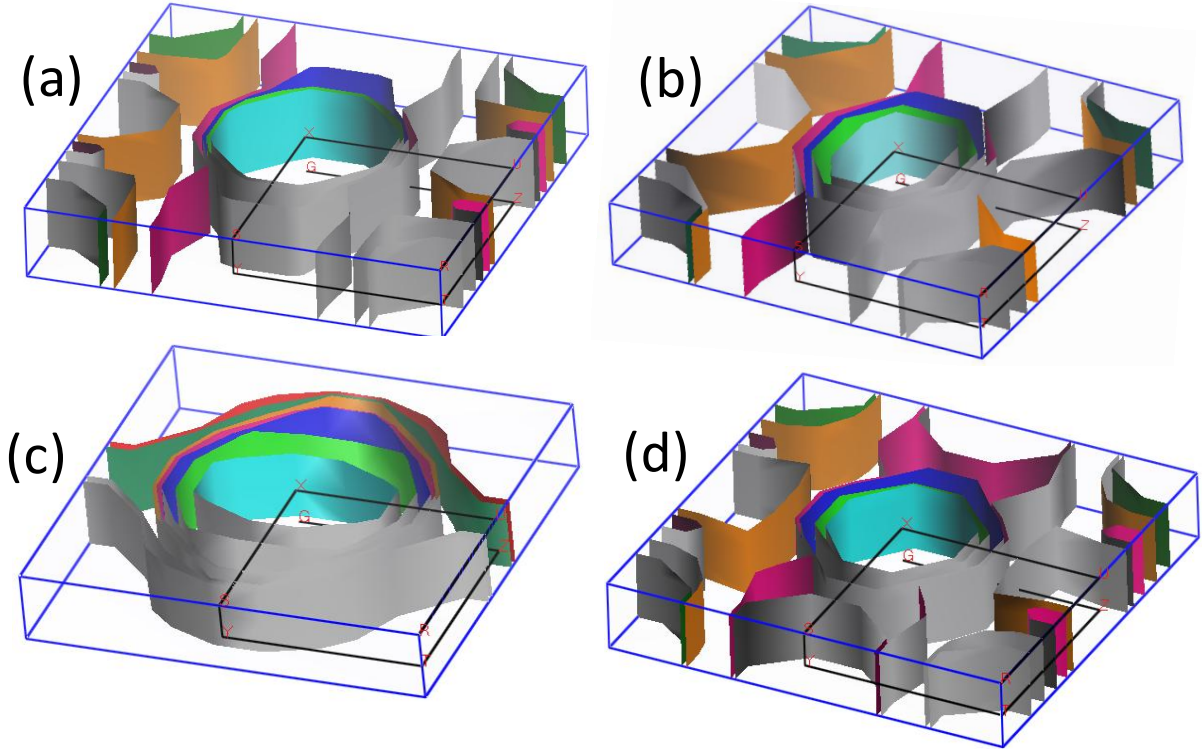

**Fig. 6.** Topology of Fermi surface of (a) $V_2BC$, (b) $Nb_2BC$,(c) $Mo_2BC$ and (d)$Ta_2BC$ compounds. The Fermi surfaces show only the bands crossing Fermi level.

### 3.4 Mechanical properties

Interests in mechanical properties of solids have been intensified because they create various opportunities for industrial applications. Many properties of the materials such as interatomic potentials, melting points, equation of states, phonon spectra etc also depend on the mechanical properties.

### 3.4.1 Elastic constants

To know the structural stability of the material, anisotropic nature of the bonding and bonding characteristic between adjacent atomic planes, the study of single crystal elastic constants are required. For the orthorhombic crystal structure, there are nine independent elastic constants. These are referred as $C_{11}$, $C_{22}$, $C_{33}$, $C_{44}$, $C_{55}$, $C_{66}$, $C_{12}$, $C_{13}$ and $C_{23}$. The calculated elastic constants along with the other theoretical values reported in the literature for $M_2BC$ (M = V, Nb, Mo and Ta) are listed in Table 2. To find a stable orthorhombic system, the elastic constants must satisfy the well known Born criteria [30]:

$C_{11} > 0;\ C_{44} > 0;\ C_{55} > 0;\ C_{66} > 0;$

$C_{11}C_{22} > C_{12}^2;\ (C_{11}C_{22}C_{33} + 2C_{12}C_{13}C_{23}) > (C_{11}C_{23}^2 + C_{22}C_{13}^2 + C_{33}C_{12}^2)$

Our calculated elastics constants $C_{ij}$ for all the compounds (Table 2) are positive and satisfy the above conditions. Therefore, the compounds under study are mechanically stable. The diagonal constants $C_{11}$, $C_{22}$ and $C_{33}$ are calculated along the directions *a*, *b* and *c*, respectively which measured the resistance to linear compression. The Mo$_2$BC compound has the largest elastic constants than that of others. It is also seen that the calculated $C_{33}$ is higher than $C_{11}$ and $C_{22}$ for all the compounds, suggesting that the incompressibility along *c* axis is stronger than that along crystallographic *a* and *b* axes. This means that the bonding strength along the [001] direction is higher than that along the [101] and [011] directions in these compounds. The off-diagonal elements ($C_{12}$, $C_{13}$, and $C_{23}$) are usually defined as shear components of elastic constants. The large differences between $C_{12}$ and $C_{13}$ are found for V$_2$BC, Mo$_2$BC and Ta$_2$BC, whereas for Nb$_2$BC, this is almost constant. This means that the compounds V$_2$BC, Mo$_2$BC and Ta$_2$BC have greater ability to resist shear along the crystallographic b and c axes under applied force than along the *a* axis. The shear component $C_{23}$ describes a uniaxial strain along *c* axis when a functional stress component is applied along the crystallographic *b* axis. The $C_{23}$ (193 GPa) for Mo$_2$BC retain the highest value, resulting small difference between $C_{22}$ (545 GPa) and $C_{33}$ (547 GPa) in magnitude. The Mo$_2$BC has also the highest value of $C_{44}$ and when the Mo atom in Mo$_2$BC is substituted by the transition metals (V, Nb and Ta), this value is gradually decreased, indicating the material's ability to resist the shear deformation in (100) plane is also decreased gradually. The study of Cauchy pressure (CP = $C_{12}$ - $C_{44}$) is important to understand the brittle/ductile behavior and the bonding nature at the atomic level of a compound [31, 32].

**Table 2:** The calculated single crystal elastic constants $C_{ij}$ (GPa) and Cauchy pressure (CP) for M$_2$BC (M = V, Nb, Mo and Ta).

| Compounds | $C_{11}$ | $C_{12}$ | $C_{13}$ | $C_{23}$ | $C_{22}$ | $C_{33}$ | $C_{44}$ | $C_{55}$ | $C_{66}$ | CP | Ref. |
|---|---|---|---|---|---|---|---|---|---|---|---|
| V$_2$BC | 474 | 148 | 159 | 115 | 450 | 525 | 161 | 257 | 146 | -13 | This |
| Nb$_2$BC | 482 | 154 | 156 | 134 | 488 | 504 | 140 | 251 | 130 | +14 | This |
| Mo$_2$BC | 520 | 209 | 218 | 193 | 545 | 547 | 176 | 265 | 180 | +33 | This |
|  | 551 | 211 | 204 | 210 | 566 | 553 | 168 | 241 | 182 | +43 | [2] |
| Ta$_2$BC | 503 | 157 | 181 | 153 | 468 | 528 | 134 | 258 | 125 | +23 | This |

A high positive value of CP signifies the increased ductile behavior of the compound. As shown in Table 2, the V$_2$BC shows a negative value of CP. On the other hand, highest positive value of CP is obtained for the Mo$_2$BC, while lower but positive values are found for the Nb$_2$BC, and Ta$_2$BC compound, disclosing the degree of ductile behavior. The negative value of CP in the V$_2$BC compound suggests that the directional covalent bonding is dominating whereas metallic bonding, govern in the Nb$_2$BC, Mo$_2$BC and Ta$_2$BC compounds due to the positive CP.

### 3.4.2 Polycrystalline elastic moduli

The bulk modulus, *B*, shear modulus, *G*, Young's modulus, *Y*, Pugh ratio, *B/G*, and Poisson ratio, *v* for M$_2$BC (M = V, Nb, Mo and Ta) are calculated from single crystal elastic constants and summarized in Table 3. The data for TiN, TiAlN, Ti$_{0.5}$Al$_{0.5}$N, Cr$_{0.5}$Al$_{0.5}$N and c-BN was listed in Table 3 for comparison. The results are compared with available experimental and theoretical data reported in literature, where available. It is found that the results are well agreed with previously reported data. The *B* and *G* for polycrystalline aggregates are estimated using well-known Voigt [34] ($B_V$, $G_V$) and Reuss [35] ($B_R$, $G_R$) approximations, being a lower and higher limit of the elastic moduli, respectively.

**Table 3 :** The calculated bulk modulus, $B$ (GPa), shear modulus, $G$ (GPa), Young's modulus, $Y$ (GPa), Pugh ratio $G/B$, and Poisson ratio, $v$ for $M_2BC$ (M = V, Nb, Mo and Ta). The data for TiN, TiAlN, $Ti_{0.5}Al_{0.5}N$, $Ti_{0.25}Al_{0.75}N$, $Cr_{0.5}Al_{0.5}N$ and c-BN was summarized for comparison only.

| Compounds | $B$ | $G$ | $Y$ | $B/G$ | $v$ | Ref. |
|---|---|---|---|---|---|---|
| $V_2BC$ | 254 | 177 | 431 | 1.44 | 0.22 | This |
|  | 260 | 178 | 435 | 1.46 |  | [6] |
| $Nb_2BC$ | 263 | 169 | 418 | 1.57 | 0.23 | This |
|  | 259 | 163 | 404 | 1.59 |  | [6] |
| $Mo_2BC$ | 317 | 187 | 469 | 1.70 | 0.25 | This |
|  | 324 | 187 | 470 | 1.73 | 0.26 | [2] |
|  | 313 | 181 | 455 | 1.73 |  | [6] |
| $Ta_2BC$ | 275 | 165 | 412 | 1.67 | 0.25 | This |
|  | 286 | 168 | 421 | 1.71 |  | [6] |
| TiN | 295 | 213 | 514 | 1.39 | 0.22 | [3] |
| TiAlN | 257 | 178 | 434 | 1.33 | 0.22 | [3] |
| $Ti_{0.5}Al_{0.5}N$ | 280 | 210 | 504* | 1.33 | 0.20* | [4] |
| $Ti_{0.25}Al_{0.75}N$ | 178 | 123 | 300* | 1.44 | 0.22* | [2] |
| $Cr_{0.5}Al_{0.5}N$ | 250 | 150* | 375 | 1.66* | 0.25 | [5] |
| c-BN | 376 | 382 | 856 | 0.98 | 0.12 | [2] |

*Calculated by us

The bulk modulus is an indicator to measure the required resistance to fracture while shear modulus measures the resistance of a material to plastic deformation under external pressure. The calculated value of $B$ (317 GPa), $G$ (187 GPa) and $Y$ (469 GPa) for $Mo_2BC$ shows the highest value than the others compound. This result indicates that the $Mo_2BC$ material is a hard and highly stiff material. The reason is that the valence electron concentration per atom of $Mo_2BC$ is much higher than that of other three compounds ($V_2BC$, $Nb_2BC$ and $Ta_2BC$). However, the calculated value of B (317 GPa of $Mo_2BC$) is 18.61 % smaller than well-known c-BN but the value (317 GPa) is 7 %, 19 %, 11.67 %, 43.84 % and 21.13 % larger than that of TiN, TiAlN, $Ti_{0.5}Al_{0.5}N$, $Ti_{0.25}Al_{0.75}N$ and $Cr_{0.5}Al_{0.5}N$, respectively. The calculated $Y$ (469 GPa) value is in very good agreement with the experimentally measured $Y$ (460 ± 21GPa) [2], which also implies the reliability of our calculations. On replacement of the Mo atom in $Mo_2BC$ compound by V, Nb and Ta elements, the value of $G$ and $Y$ are found to be decreased in the sequence of $G$ ($V_2BC$) > $G$ ($Nb_2BC$) > $G$ ($Ta_2BC$) and $Y$ ($V_2BC$) > $Y$ ($Nb_2BC$) > $Y$ ($Ta_2BC$), while the value of $B$ is seen to increase in the sequence $B$ ($V_2BC$) < $B$ ($Nb_2BC$) < $B$ ($Ta_2BC$), indicating that $Ta_2BC$ is more stiffer than the $V_2BC$ and $Nb_2BC$. It is interesting to note here that even the value of $B$

(275 GPa) for $Ta_2BC$ is 35.27 % higher than that of benchmark cutting tool material, $Ti_{0.25}Al_{0.75}N$. Pugh [35] proposed a famous modulus ratio between $G$ and $B$, known as Pugh's ratio which separates the failure mode (ductile and brittleness) of a materials. The Poisson's ratio ($\upsilon$) is also an indicator to separate between brittle and ductile materials [36]. If the value of $B/G$ and $\upsilon$ is larger (smaller) than 1.75 and 0.26 for a material then it is said to be brittle (ductile), respectively [28]. According to these criteria, $V_2BC$ and $Nb_2BC$ are brittle in nature. It may be noted that the $Mo_2BC$ and $Ta_2BC$ compounds are observed not so far from the brittle-ductile transition line. The value of CP and $B/G$ (Table 2 and 3) reveals these two materials, therefore, may be considered as moderately ductile material compared to other typical hard coating materials summarized in Table 3. Furthermore, the value of $\upsilon$ also assists to predict the bonding nature in the material. The value of $v$ is typically 0.10 for covalent bonding while it is 0.33 for metallic bonding in materials [37]. The Poisson's ratio of these compounds lies between two characteristic values. Therefore, the chemical bonding should be a mixture of ionic and metallic in nature. Also, the Poisson's ratio gives more information regarding the character of the bonding forces than any of the other elastic coefficients. As the reported range of $v$ for central-force solids is 0.25−0.50, otherwise it is non-central force solid [38], thus, the value of $v$ for $Mo_2BC$ and $Ta_2BC$ compounds suggests that the central inter-atomic forces are involved whereas for $V_2BC$ and $Nb_2BC$ indicates non-central force in these solids.

## 3.5 Structural anisotropy

Most of the known crystals are usually elastically anisotropic. The study of elastic anisotropy of a material is a matter of deep interest both in crystal physics and industrial fields as anisotropy induces many mechanical-physical properties including unusual phonon modes, precipitation, internal friction, phase transformations, dislocation dynamics etc.

For an orthorhombic system, the elastic anisotropy can be calculated in different modes and the relevant formulae can be found elsewhere [28,39-42]. The calculated anisotropic factors are listed in Table 4. It is observed that the compounds considered here are anisotropic in nature, because the value $A_i = 1$, where i = 1, 2 and 3 represents complete elastic isotropy, while values smaller or greater than 1 measure the degree of elastic anisotropy.

**Table 4:** Shear anisotropic factors ($A_i$, where i = 1, 2 and 3) and bulk modulus (in GPa) $B_a$, $B_b$ and $B_c$ along crystallographic axes $a$, $b$ and $c$, respectively. The anisotropy of linear bulk moduli ($A_{Ba}$ and $A_{Bc}$) along $a$ and $c$ axes, anisotropy factors in compressibility ($A_B$) and shear moduli ($A_G$) in percentage, and universal anisotropic index, $A^U$ for M$_2$BC (M = V, Nb, Mo and Ta) compounds.

| Compound | $A_1$ | $A_2$ | $A_3$ | $B_a$ | $B_b$ | $B_c$ | $A_{Ba}$ | $A_{Bc}$ | $A_B$ | $A_G$ | $A^U$ |
|---|---|---|---|---|---|---|---|---|---|---|---|
| V$_2$BC | 0.94 | 1.38 | 0.93 | 809 | 663 | 833 | 1.22 | 1.25 | 0.251 | 2.134 | 0.223 |
| Nb$_2$BC | 0.83 | 1.38 | 0.78 | 799 | 763 | 801 | 1.04 | 1.05 | 0.012 | 2.622 | 0.269 |
| Mo$_2$BC | 1.11 | 1.50 | 1.11 | 939 | 942 | 972 | 0.99 | 1.03 | 0.005 | 1.672 | 0.170 |
| Ta$_2$BC | 0.80 | 1.50 | 0.76 | 861 | 719 | 915 | 1.19 | 1.27 | 0.227 | 3.247 | 0.340 |

The value of bulk modulus (in GPa) $B_a$, $B_b$ and $B_c$ along crystallographic axes $a$, $b$ and $c$, respectively indicates the bulk modulus along $c$ axis is much significant than in other directions. From the anisotropy of directional bulk modulus ($A_{Ba}$ and $A_{Bc}$) values, it gives the further confirmation of elastic anisotropy. The concept of percent elastic anisotropy in polycrystalline solids for non-cubic system was reported [40]. The percentage anisotropy in compressibility ($A_B$) and shear ($A_G$) as well as universal anisotropic index ($A^U$) can be calculated by the following equations:

$A_B = \frac{B_V - B_R}{B_V + B_R} \times 100\%$, $A_G = \frac{G_V - G_R}{G_V + G_R} \times 100\%$ and $A^U = 5\frac{G_V}{G_R} + \frac{B_V}{B_R} - 6 \geq 0$ where the subscripts, $V$ and $R$, correspond to the Voigt and Reuss bounds, respectively. When the values of $A_B$, $A_G$ or $A^U$ are equal to zero, the material is completely isotropic and the deviation from zero defines the extent of anisotropy associated with crystals. It is interesting to note that the level of anisotropy is much higher in shear than in compressibility for the compounds considered here. From the data shown in Table 4, it can be concluded that the studied compounds are mechanically anisotropic in nature.

### 3.6 Mulliken population and Vicker hardness

Hardness of a polycrystalline material can be defined as the resistance offered by a given material to external mechanical action and is of utmost interest to enhance the knowledge about a material in the heavy duty industrial applications. Using first principles technique, Mulliken population analysis helps us to explain the distribution of electron density in various bonds. The study of overlap population gives quantitative figures which can be taken as measures of bonding and anti-bonding strengths. If the value of total overlap population for the nearest neighbors in the crystal is positive, they are bonded; however, if negative, they are anti-bonded. The value of overlap population is close to zero means no significant interaction between the electronic populations of the two atoms and can be ignored to calculate the hardness of the material. A high overlap can also indicate a high degree of covalency in the bond. The Mulliken charge of a particular atomic species and the overlap population between two atoms can be calculated by established formulae [40,43,44]. The effective valence charge (EVC) is defined as the difference between the formal ionic charge and the Mulliken charge on the anion species in the crystal. The nature of the chemical bonding in a material can be identified from this charge. A zero value of EVC indicates an ideal ionic bond while the values of EVC greater than zero suggest increasing levels of covalency.

Table 5 shows Mulliken atomic and bond overlap population for nearest neighbors of $M_2BC$ (M = V, Nb, Mo and Ta) compounds. It is seen that there is no bonding between C-C atoms as the value of bond overlap population is negative. A bond between C-V/Nb/Mo/Ta indicates the highest degree of covalency compared to B-V/Nb/Mo/Ta. Also, the B-B bond in the compounds under study shows the maximum degree of covalency. The analysis of atomic populations also helps us to understand the charge transfer mechanism from one atom to another. It is realized that in $V_2BC$ compound the transfer of charge from V to C and B is 0.62 and 0.56e, respectively. Similar charge transfer mechanism can also be realized from other compounds under investigation.

Understanding the hardness of any materials is very important for practical device applications. The correlation between elastic polycrystalline module and hardness provide a deep understanding for the mechanical behavior. Gou et al. [45] found a correct relation to calculate the Vickers hardness for metallic compounds. The hardness of a multiband material can be calculated from the geometrical average of all individual bond hardness by the following relation [40,46]:

$H_V = [\Pi^\mu (H_v^\mu)^{n^\mu}]^{1/\Sigma n^\mu}$ ,where $n^\mu$ refers the number of bond of type $\mu$ composing the real multiband crystals.

The hardness and Mulliken bond overlap population of $M_2BC$ (M = V, Nb, Mo and Ta) compounds are listed in Table 6. The values of Vickers hardness of $M_2BC$ (M = V, Nb, Mo and Ta) are 10.71, 12.44, 8.52 and 16.80 GPa, respectively. The Vickers hardness increases by the following the sequence of M-element in $M_2BC$ compounds: Mo to V to Nb to Ta. In other words, the hardness decreases when the M element moves up from bottom of the group in the periodic table. The $Mo_2BC$ compound is expected to be reasonable soft and machinable. The $V_2BC$, $Nb_2BC$ and $Ta_2BC$ compounds are not easily machinable for conventional cutting tools due to comparatively high hardness value. The hardness values ranging from 8.52-16.80 GPa of these compounds are higher than that (2-8 GPa) of many well-known MAX phase compound [24, 29, 46].

**Table 5:** Mulliken atomic and bond overlap population of $M_2BC$ (M = V, Nb, Mo and Ta) compounds.

| Compound | Atoms | Mulliken atomic population | | | | | | Mulliken bond overlap population | | | |
|---|---|---|---|---|---|---|---|---|---|---|---|
| | | s | p | d | Total | Charge (e) | EVC (e) | Bond | Bond number $n^\mu$ | Bond length $d^\mu$ (Å) | Bond overlap population $P^\mu$ |
| $V_2BC$ | C | 1.45 | 3.17 | 0.00 | 4.62 | -0.62 | --- | C-V | 4 | 1.96 | 0.19 |
| | B | 0.98 | 2.58 | 0.00 | 3.56 | -0.56 | --- | C-V | 4 | 1.99 | 0.22 |
| | V | 2.04 | 6.51 | 3.81 | 12.36 | 0.64 | 4.36 | C-V | 4 | 2.10 | 1.41 |
| | V | 2.14 | 6.57 | 3.74 | 12.45 | 0.55 | 4.45 | B-V | 4 | 2.22 | 0.75 |
| | | | | | | | | B-V | 4 | 2.70 | 0.03 |
| | | | | | | | | V-V | 4 | 2.89 | 0.04 |
| | | | | | | | | B-B | 2 | 1.76 | 1.42 |
| | | | | | | | | C-C | 2 | 2.92 | -0.14 |
| $Nb_2BC$ | C | 1.45 | 3.26 | 0.00 | 4.71 | -0.71 | --- | C-Nb | 4 | 2.14 | 0.25 |
| | B | 0.98 | 2.56 | 0.00 | 3.54 | -0.54 | --- | C-Nb | 4 | 2.16 | 0.40 |
| | Nb | 2.08 | 6.34 | 3.93 | 12.35 | 0.65 | 4.35 | C-Nb | 4 | 2.25 | 1.34 |
| | Nb | 2.17 | 6.43 | 3.80 | 12.40 | 0.60 | 4.40 | B-Nb | 4 | 2.39 | 0.09 |
| | | | | | | | | B-Nb | 4 | 2.41 | 0.81 |
| | | | | | | | | Nb-Nb | 4 | 3.15 | 0.17 |
| | | | | | | | | B-B | 2 | 1.82 | 1.44 |
| | | | | | | | | C-C | 2 | 3.16 | -0.10 |
| $Mo_2BC$ | C | 1.43 | 3.19 | 0.00 | 4.61 | -0.61 | --- | C-Mo | 4 | 2.08 | 0.18 |
| | B | 0.95 | 2.51 | 0.00 | 3.46 | -0.46 | --- | C-Mo | 4 | 2.15 | 0.22 |
| | Mo | 2.06 | 6.44 | 4.99 | 13.49 | 0.51 | 5.49 | C-Mo | 4 | 2.18 | 1.38 |
| | Mo | 2.16 | 6.45 | 4.83 | 13.43 | 0.57 | 5.43 | C-Mo | 4 | 4.67 | 0.04 |
| | | | | | | | | B-Mo | 4 | 2.31 | 0.79 |
| | | | | | | | | B-B | 2 | 1.82 | 1.34 |
| | | | | | | | | Mo-Mo | 4 | 2.95 | -0.22 |
| | | | | | | | | C-C | 2 | 2.99 | -0.12 |
| $Ta_2BC$ | C | 1.50 | 3.25 | 0.00 | 4.74 | -0.74 | --- | C-Ta | 4 | 2.17 | 0.36 |
| | B | 1.08 | 2.55 | 0.00 | 3.63 | -0.63 | --- | C-Ta | 4 | 2.19 | 0.43 |
| | Ta | 0.32 | 0.13 | 3.86 | 4.31 | 0.69 | 4.31 | C-Ta | 4 | 2.28 | 1.58 |
| | Ta | 0.41 | 0.17 | 3.74 | 04.32 | 0.68 | 4.32 | B-Ta | 4 | 2.43 | 1.14 |
| | | | | | | | | B-Ta | 4 | 3.07 | 0.14 |
| | | | | | | | | Ta-Ta | 4 | 3.17 | 0.38 |
| | | | | | | | | Ta-Ta | 2 | 3.14 | 0.22 |
| | | | | | | | | B-B | 2 | 1.85 | 1.34 |
| | | | | | | | | C-C | 2 | 3.19 | -0.09 |

**Table 6:** Muliken bond overlap population of $\mu$-type bond $P^{\mu}$, bond length $d^{\mu}$, metallic population $P^{\mu'}$, bond volume $v_b^{\mu}$, Vickers hardness of $\mu$-type bond $H_V^{\mu}$ and $H_V$ of $M_2BC$ (M = V, Nb, Mo and Ta) compounds.

| Compounds | Bond | $d^{\mu}$ | $P^{\mu}$ | $P^{\mu'}$ | $v_b^{\mu}$ | $H_V^{\mu}$ | $H_V$ |
|---|---|---|---|---|---|---|---|
| $V_2BC$ | C-V | 1.960 | 0.19 | | 0.037 | 4.207 | |
| | C-V | 1.990 | 0.22 | | 0.0341 | 4.598 | |
| | C-V | 2.100 | 1.41 | 0.038 | 0.026 | 26.466 | 10.71 |
| | B-V | 2.220 | 0.75 | | 0.020 | 10.498 | |
| | B-B | 1.760 | 1.42 | | 0.063 | 64.284 | |
| $Nb_2BC$ | C-Nb | 2.139 | 0.25 | | 0.055 | 9.380 | |
| | C-Nb | 2.160 | 0.40 | | 0.053 | 14.770 | |
| | C-Nb | 2.251 | 1.34 | 0.021 | 0.043 | 41.750 | 12.44 |
| | B-Nb | 2.407 | 0.81 | | 0.031 | 17.870 | |
| | Nb-Nb | 3.150 | 0.17 | | 0.008 | 0.880 | |
| | B-B | 1.814 | 1.44 | | 0.126 | 132.160 | |
| $Mo_2BC$ | C-Mo | 2.082 | 0.18 | | 0.029 | 2.960 | |
| | C-Mo | 2.151 | 0.22 | | 0.025 | 3.244 | |
| | C-Mo | 2.182 | 1.38 | 0.042 | 0.023 | 22.658 | 8.52 |
| | B-Mo | 2.310 | 0.79 | | 0.017 | 9.526 | |
| | B-B | 1.816 | 1.34 | | 0.057 | 55.083 | |
| $Ta_2BC$ | C-Ta | 2.170 | 0.36 | | 0.105 | 24.805 | |
| | C-Ta | 2.198 | 0.43 | | 0.098 | 28.365 | |
| | C-Ta | 2.280 | 1.58 | 0.041 | 0.082 | 93.387 | 16.80 |
| | B-Ta | 2.434 | 1.14 | | 0.059 | 48.101 | |
| | B-Ta | 3.067 | 0.14 | | 0.019 | 1.368 | |
| | Ta-Ta | 3.179 | 0.38 | | 0.016 | 3.906 | |
| | Ta-Ta | 3.136 | 0.22 | | 0.017 | 2.210 | |
| | B-B | 1.845 | 1.34 | | 0.236 | 227.178 | |

### 3.7 Optical properties

Different optical functions such as photo-conductivity, absorption, reflectivity, loss function, refractive index and dielectric function of the materials are thoroughly discussed in this section. The interaction of a photon with the electrons in the system is described in terms of time dependent perturbations of the ground state electronic states. Transitions between occupied and unoccupied states are caused by the electric field of the photon. Therefore, the study of the optical properties is very useful to examine the electronic character of that material. The optical properties are usually calculated from the complex dielectric function, where $\varepsilon_1$ is the real part of dielectric function derived from the Kramer-Kronig relations and $\varepsilon_2$ is the imaginary part of the dielectric function which is computed from the momentum matrix elements between the occupied and unoccupied wave functions. The equations for calculations of different optical functions can be found elsewhere [47, 48].

A smearing value of 0.5 eV is used to specify Gaussian broadening to be applied for calculating the optical functions of the compounds. The Drude plasma frequency of 3 eV and damping factor of 0.05 eV are included due to the metallic crystal structure which is confirmed from the band structure calculations (Fig. 2).

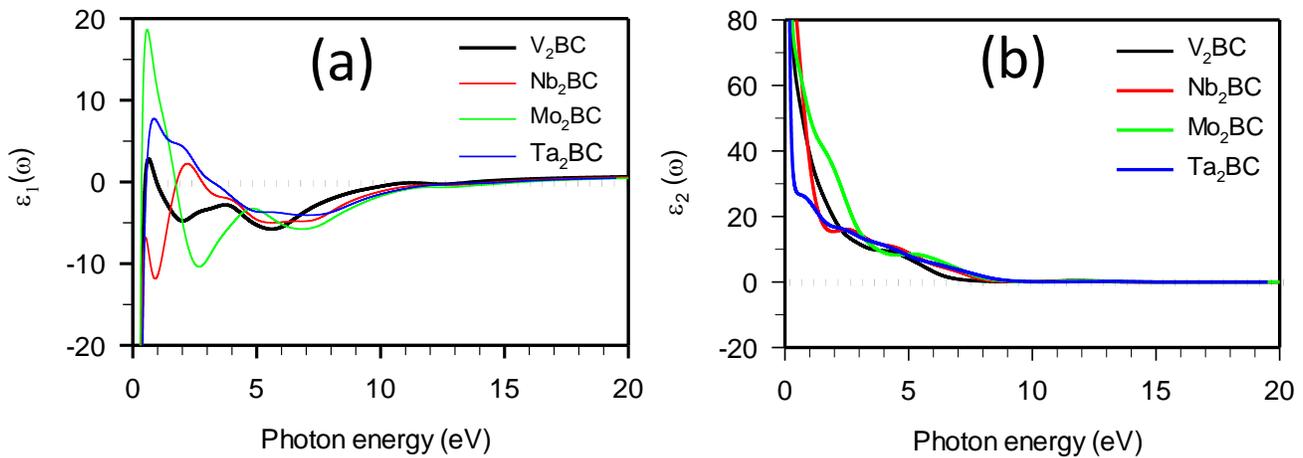

**Fig. 7.** Dielectric functions: (a) real part ($\varepsilon_1$) and (b) imaginary part ($\varepsilon_2$) for $V_2BC$, $Nb_2BC$, $Mo_2BC$ and $Ta_2BC$ compounds as a function of photon energy.

The real part $\varepsilon_1$ of dielectric function for the $V_2BC$, $Nb_2BC$, $Mo_2BC$ and $Ta_2BC$ compounds are shown in Fig. 7(a). The intraband contribution to the optical properties plays an important role and affects mainly the low energy infrared part of the spectra [49]. All peaks around less than 1 eV are assigned due to intraband electronic transitions of the bands. The large negative values of $\varepsilon_1$ indicate that the materials have a drude-like behavior whereas at higher energies the interband transitions occur giving rise to optical features. The spectra exhibit different optical feature in the energy ranges 1-5 eV and at higher energies, almost similar features are observed. The imaginary part $\varepsilon_2$ of the dielectric function are depicted in Fig. 7(b). It is observed that the values of $\varepsilon_2$ go through zero from above at around 9 eV. This is a further indication of metallic behavior of the compounds.

The refractive indexes and the extinction coefficient of the compounds are presented in Fig.8 (a) and (b) The values of static refractive index, $n(0)$ of the $V_2BC$, $Nb_2BC$, $Mo_2BC$ and $Ta_2BC$ are found to be 2.1, 2.5, 1.9 and 0.8, and the main peaks are at around 1, 1.2, 3 and 1.5 eV photon energy, respectively. All these peaks approach zero at the ranges around ~ 14-15 eV photon energy. The extinction coefficient $k$ spectra of the compounds exhibit almost similar feature and go through zero from above at around 9 eV.

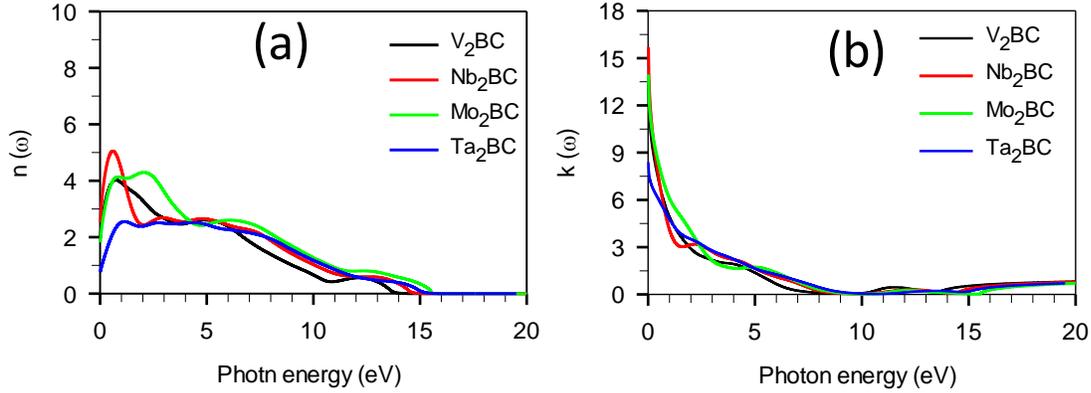

**Fig. 8.** (a) Refractive index (n) and (b) extinction coefficient (k) for $V_2BC$, $Nb_2BC$, $Mo_2BC$ and $Ta_2BC$ compounds as a function of photon energy.

Fig. 9 (a) exhibits the absorption coefficients. The absorption coefficient indicates the fraction of energy lost by the light wave when it penetrates through the material. Some distinct features of absorption spectra are clearly visualized although for the compounds, these spectra almost follow the same route. The absorption spectra rise sharply and then achieve maximum peak. Finally these spectra decrease vigorously with some intermediate peaks in the region of 12-13 eV. The highest absorption spectra (HAS) for $Mo_2BC$ compound is found at 7.5 eV while for $V_2BC$, $Nb_2BC$ and $Ta_2BC$ compounds, the HAS are at 5.8, 7.2 and 7.8 eV, respectively. This result indicates that when the Mo of $Mo_2BC$ is replaced by V, Nb and Ta, the HAS peaks are gradually shifted with photon energy. In metallic compounds, the absorption spectra start when the photon energy is zero as shown in Fig. 9(a).

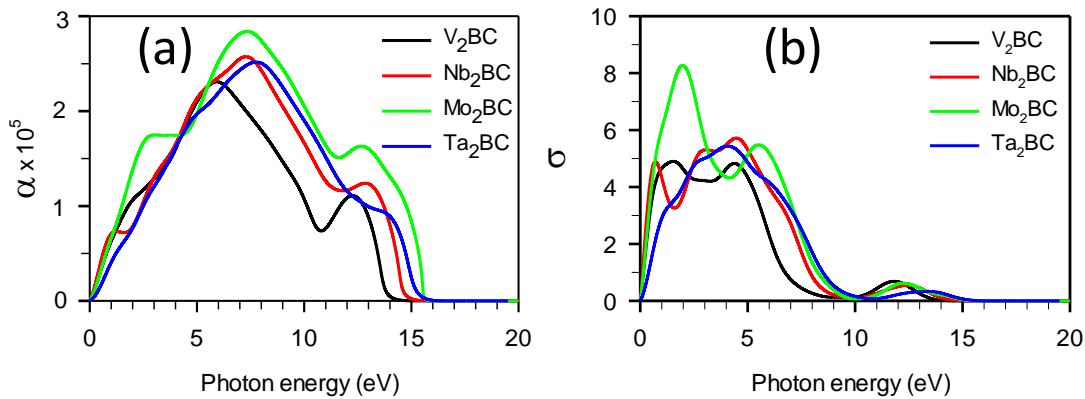

**Fig. 9.** (a) Absorption coefficient ($\alpha$) and (b) photo-conductivity ($\sigma$) for $V_2BC$, $Nb_2BC$, $Mo_2BC$ and $Ta_2BC$ compounds as a function of photon energy.

Fig. 9(b) shows the calculated real part of the photo-conductivity for different materials as a function of incident photon energy. Photo-conductivity shows metallic character [50–52]. Optical conductivity spectra clearly exhibit different maxima for the studied compounds. The $Mo_2BC$ is the highly photo-conductive when the incident photon energy is around 2 eV. The highest peaks (maximum value of the conductivity) on the spectra are gradually shifted (except for $Ta_2BC$) with increasing incident photon energy. The maximum values are found at 1.5, 4.8 and 4.0 eV photon energies for the $V_2BC$, $Nb_2BC$ and $Ta_2BC$ compounds, respectively. With further increase of photon energy, photo-conductivity gradually decreases and finally goes to zero at around 9-10 eV.

Another important property that can be calculated from the complex dielectric constant is the energy loss function as shown in Fig. 10(a). It describes the energy loss by an electron passing through a material. There is no loss spectrum for any compounds in the region up to the photon energy of 10 eV. A loss feature is observed in the range of 10 -16 eV. In these energy ranges the peaks of the energy loss spectrum could be very large while the value of $\varepsilon_2 < 1$ and $\varepsilon_1 = 0$. This large peak is associated with the so called bulk plasma frequency of the material. The bulk plasma frequencies of $V_2BC$, $Nb_2BC$, $Ta_2BC$ and $Mo_2BC$ compounds are found to be 13.7, 14.5, 15 and 15.6 eV, respectively, which will indicate to the rapid diminish of the reflectivity as shown in Fig.10(b). After this frequency, the energy loss spectra could not exhibit any distinct maxima as $\varepsilon_2$ approaches to zero.

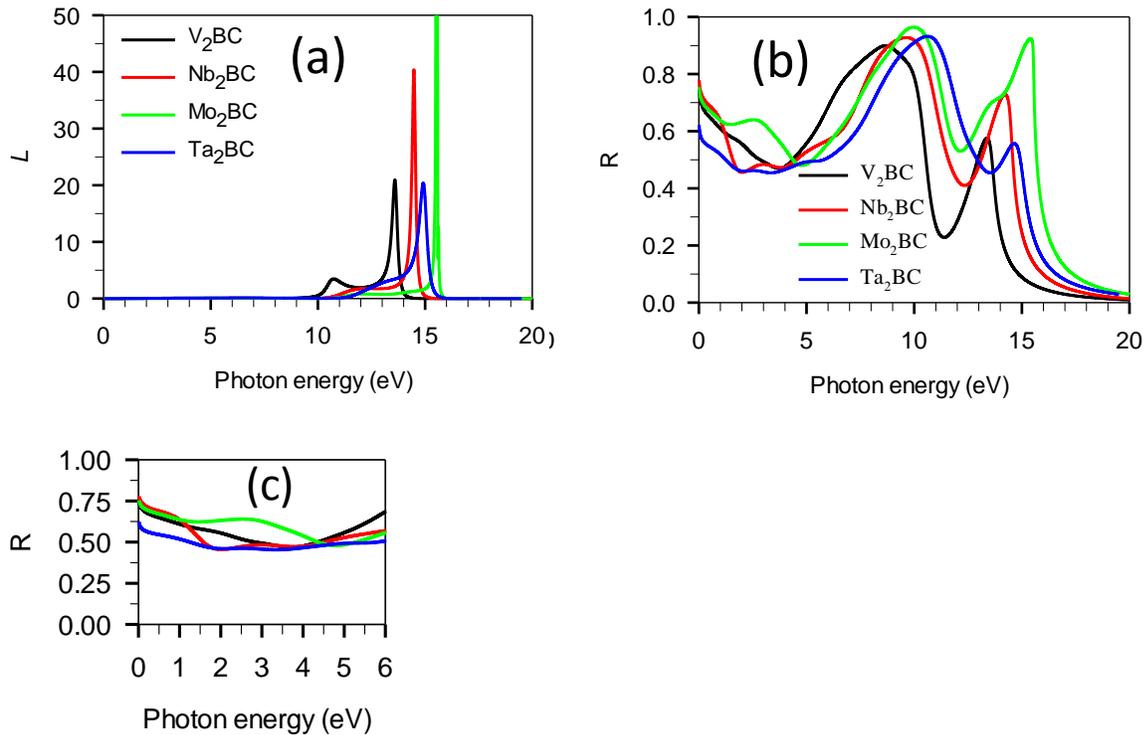

**Fig. 10.** (a) Loss function (L) and (b) Reflectivity (R) for $V_2BC$, $Nb_2BC$, $Mo_2BC$ and $Ta_2BC$ compounds as a function of photon energy. Fig. (C) also shows reflectivity (R) in the energy range upto 6 eV photon energy.

The reflectivity spectra as a function of photon energy are depicted in Fig. 10(b). It is seen from Fig.10 (c) that the reflectivity spectra have no significant change in the near infrared, visible light and near ultraviolet region (upto ~ 6 eV). In this energy range, the values of reflectivity spectra of the $V_2BC$, $Nb_2BC$ and $Mo_2BC$ compounds are always higher than that of $Ta_2BC$. Solar heating will be reduced if a material has reflectivity around 44% in the visible light region as reported by Li et al. [53]. The reflectivity never drops below 50%, which is an indication that these materials could be used as a promising coating material to reduce solar heating. It is noted here that the value of reflectivity spectra of the compounds under investigation are higher than that of some of the MAX compounds [29, 46, 54, 55]. Some sharp peaks of reflectivity spectra are found in the energy range 9-11 eV and 14-16 eV and finally reflectivity approaches zero around 20 eV. In other words, the highest values (95-97%) of the reflectivity spectra are found in the ultraviolet region. (Do not use the term 'so called MAX compounds')

## 3.8 Debye temperature, thermal conductivity and Grüneisen parameter

Debye characteristic temperature, $\Theta_D$, a fundamental key parameter of solids, plays an important role on melting temperature, specific heats, thermal expansion, lattice vibration and thermal conductivity. The low and high temperature region for a solid can be usually distinguished by the study of $\Theta_D$. Debye temperature can be calculated from knowledge of elastic constants via the calculation of average sound velocity using the following equation[56]:

$$\Theta_D = h/k_B \left[(3n/4\pi)N_A\rho/M\right]^{1/3} v_m,$$

where M is the molar mass, n is the number of atoms in the molecules, and $\rho$ is the mass density ; h is the Plank's constant, $k_B$ is the Boltzmann constant, $N_A$ is the Avogadro's number and $v_m$ is the average sound velocity. The $v_m$ can be determined from the longitudinal ($v_l$) and transverse ($v_t$) sound velocities in an isotropic material using the following expression:

$$v_m = \left[1/3\left(1/v_l^3 + 2/v_t^3\right)\right]^{-1/3}$$

The $v_l$ and $v_t$ can be expressed in terms of polycrystalline bulk modulus, B and shear modulus, G from the relation derived by Voigt and Reuss, by the relations:

$$v_l = [(3B+4G)/3\rho]^{1/2} \text{ and } v_m = [G/\rho]^{1/2}.$$

The calculated $\Theta_D$ of $M_2BC$ (M = V, Nb, Mo and Ta) compounds are listed in Table 7. A gradually decreasing trend in $\Theta_D$ is realized when Mo atom is replaced by the same group elements (V, Nb and Ta) in the periodic table. A higher $\Theta_D$ usually corresponds to a higher phonon thermal conductivity. Therefore, $V_2BC$ should be more thermally conductive than the other compounds. The minimum thermal conductivity, $K_{min}$ and Grüneisen parameter, $\gamma$ can be obtained by the equation reported in [28, 57]. The value $K_{min}$ is found to be in the following order: $K_{min}$ ($V_2BC$) > $K_{min}$ ($Mo_2BC$) > $K_{min}$ ($Nb_2BC$) > $K_{min}$ ($Ta_2BC$). $\gamma$ measures the anharmonic effect in crystalline solids and the values (1.36, 1.40. 1.50 and 1.51) obtained gradually increase with M = V, Nb, Mo and Ta, respectively. It is noteworthy that $Ta_2BC$ has the lowest value of $\Theta_D$ (508 K) and $K_{min}$ (1.03 $Wm^{-1}K^{-1}$). These values for $Ta_2BC$ are much lower than that of well-known thermal barrier coating (TBC) material, $Y_4Al_2O_9$ [57]. $Ta_2BC$, therefore, can be used as promising TBC material.

**Table 7:** Calculated density, longitudinal, transverse and average sound velocities ($v_l$, $v_t$, and $v_m$), Debye temperature, $\Theta_D$, minimum thermal conductivity, $K_{min}$ and Grüneisen parameter, $\gamma$ for $M_2BC$ (M = V, Nb, Mo and Ta)

| Compounds | $\rho$ (gm/cm$^3$) | $v_l$(m/s) | $v_t$(m/s) | $v_m$(m/s) | $\Theta_D$ (K) | $K_{min}$(W/(mK) | $\gamma$ |
|---|---|---|---|---|---|---|---|
| V$_2$BC | 5.63 | 9331 | 5608 | 6202 | 881 | 1.95 | 1.36 |
| Nb$_2$BC | 7.50 | 8068 | 4746 | 5260 | 693 | 1.42 | 1.40 |
| Mo$_2$BC | 8.60 | 8116 | 4664 | 5180 | 707 | 1.50 | 1.50 |
| Ta$_2$BC | 13.33 | 6094 | 3518 | 3906 | 508 | 1.03 | 1.51 |

## 4. Conclusions

The theoretical calculations based on density functional theory have been performed on a series of metallic boro-carbides, $M_2BC$ (M = V, Nb, Mo and Ta) to study the structural, elastic, optical, thermodynamic and electronic properties along with charge density and Fermi surface topology. Very good agreements of the calculated structural parameters and elastic constants with available experimental and theoretical results confirm the reliability of our computations. The valance band and conduction band appreciably overlap with each other for all the compounds; as a consequence there is no band gap at the Fermi level, indicating metallic behavior. The Fermi surfaces topologies for the V$_2$BC, Nb$_2$BC, Mo$_2$BC and Ta$_2$BC are formed mainly by the low-dispersive *4d/5p* and *2p* orbitals, which are primarily responsible for the electrical conductivity of the compounds. The analysis of atomic populations reveals that the B-B bond in all the materials under study shows the maximum degree of covalency compared to other bonds such as C-V/Nb/Mo/Ta andB-V/Nb/Mo/Ta. The central inter-atomic forces are involved in Mo$_2$BC and Ta$_2$BC compounds while non-central forces are involved in V$_2$BC and Nb$_2$BC compounds. The results of anisotropic bulk modulus reveal that the bulk modulus along *c* axis is much significant than in other directions for the compounds considered here. Moreover, the level of anisotropy is much higher in shear than in compressibility for the compounds. The values of Vickers hardness of $M_2BC$ (M = V, Nb, Mo and Ta) are found to be 10.71, 12.44, 8.52 and 16.80 GPa, respectively, implying that hardness decreases when the Mo element is replaced by the same group elements that move up from the bottom in the periodic table. The hardness of $M_2BC$ exceeds that for many well known MAX phase compounds. The bulk plasma frequencies of $M_2BC$ (M = V, Nb, Mo and Ta) are found to be 13.7, 14.5, 15 and 15.6 eV, respectively. The transition from metallic to dielectric behavior starts at these bulk plasma frequencies. A gradually decreasing trend in the Debye temperature ($\Theta_D$) and thermal conductivity ($K_{min}$) is found when Mo atom is replaced by the same group elements (V, Nb and Ta) in the periodic table. The values of $K_{min}$(1.03 Wm$^{-1}$K$^{-1}$) and $\Theta_D$(508 K) of Ta$_2$BC are lower than that (1.13 Wm$^{-1}$K$^{-1}$ and 564 K) of a typical and widely reported TBC material, Y$_4$Al$_2$O$_9$, indicating that Ta$_2$BC compound is appealing for a TBC material. All the metallic boro-carbides studied here show attractive reflectivity characteristics which indicate that M$_2$BC are potential coating materials capable of reducing solar heating.